\documentclass[runningheads]{llncs}

\usepackage{graphicx}
\usepackage{epsfig} 
\usepackage{mathptmx} 
\usepackage{times}
\usepackage{amsmath} 
\usepackage{amssymb} 
\usepackage{abbrevs}
\usepackage{booktabs}
\usepackage{mdframed}

\begin{document}

\title{
Asymptotic Performance Analysis of Blockchain Protocols
}

\author{
Antoine Durand\inst{1}\inst{2} 
\and
Elyes Ben-Hamida\inst{1}
\and
David Leporini\inst{4}
\and
G\'erard Memmi\inst{3}
}
\institute{
IRT SystemX, Paris-Saclay, France
\email{firstname.lastname@irt-systemx.fr}
\and
Universit\'e Paris-Saclay, France
\and
LTCI, Telecom ParisTech
\email{firstname.lastname@telecom-paristech.fr}
\and
Atos, Les Clayes-sous-BoisŸ
\email{firstname.lastname@atos.net}
}

\maketitle

%%%%%%%%%%%%%%%%%%%%%%%%%%%%%%%%%%%%%%%%%%%%%%%%%%%%%%%%%%%%%%%%%%%%%%%%%%%%%%%%

\begin{abstract}

In the light of the recent fame of Blockchain technologies, numerous proposals and projects aiming at better practical viability have emerged. However, formally assessing their particularities and benefits has proven to be a difficult task.
The aim of this work is to compare the fundamental differences of such protocols to understand how they lead to different practical performances.
To reach this goal, we undertake a complexity analysis of a wide range of prominent distributed algorithms proposed for blockchain systems, under the lens of Total Order Broadcast protocols.
We sampled protocols designed for very different settings and that use a broad range of techniques, thus giving a good overview of the achievements of state-of-the-art techniques. 
By analyzing latency and network usage, we are able to discuss each protocol's characteristics and properties in a consistent manner. 
One corollary result to our work is a more robust criteria to classify protocols as permissioned or permissionless.

\end{abstract}

\section{Introduction}

Nakamoto's famous work on Bitcoin~\cite{BTC} showed the potential for applications of Byzantine agreement protocols. One of Bitcoin's main achievements is being efficient at large scale, such that the protocol stays practical even in the so-called permissionless setting, \ie, when participation is open to anyone and not restricted. 
This contrasts with the existing line of research for fault tolerant distributed systems that were mostly suited for small-scale systems. 
Since Bitcoin, numerous propositions and publications from both scholars and industrials emerged, ranging from proofs of concept, small improvements to new full-fledged protocols supporting various environments such as large data center or network of smart objects (aka Internet of Things) ~\cite{HQGM}. Many of these propositions are targeted toward improving performance in terms of confirmation time and transactions per seconds. Indeed, these metrics are directly visible by the end users, and Bitcoin is notoriously weak in those regards compared to others existing protocols such as PBFT~\cite{PBFT}.

\subsection{Motivation}

Although progress has been made to understand how Bitcoin-like protocols relates to existing research~\cite{FormalOverview,BTCAcademic}, the contribution and relevance of each proposition is difficult to compare and assess. Additionally, considering the highly competitive climate of cryptocurrencies, the good faith of each authors' description and evaluation may be questioned.

As a result, it is difficult for the users to draw a coherent view of the blockchain ecosystem without making an exhaustive technological deep dive. Worse, each project's practical viability may greatly vary even across similar protocols, due to different implementation, usage and context. 

In order to shed some light on this issue, we start from the simple observation that most blockchain protocols aim at implementing some variant of Total Order Broadcast (TO Broadcast), also referred to as Atomic Broadcast.
Informally, a TO Broadcast is a protocol that allows processes to reliably broadcast messages, such that messages order are agreed on by the whole network. Hence, we can readily use TO Broadcast as an abstraction layer of the functionality offered by blockchain protocols, which allows for a generic discussion on their properties. In particular, it will serve as a basis to define common performance metrics.

This work will evaluate the efficiency of prominent blockchain protocols, when seen as a way to implement TO broadcast. 
The aim of this paper is to compare each protocol efficiency related to its network usage. As such, we will evaluate latency and communication complexity as functions of network parameters. To better relate them with Bitcoin, we will particularly focus on their scalability in terms of number of participants. 
This helps us to better understand the different trade-offs adopted by existing algorithms to reach their goals in terms of performance metrics that are relevant to the end user. 

\subsection{Paper results} 

We have analyzed a wide range of algorithms: Bitcoin~\cite{BTC,GKL17}, Snow White~\cite{SnowWhite}, Spectre~\cite{Spectre}, Ouroboros~\cite{Ouroboros}, Algorand~\cite{Algorand,AlgorandBA}, Tendermint~\cite{tendermintthesis} and HoneyBadgerBFT~\cite{HBBFT}.

In this paper we were able to compare each algorithm through a generic framework, by instantiating a few parameters describing each assumed model.
Thus, these protocols can be consistently analyzed and compared in terms of communication complexity (overall number of bits sent through the network) and latency (the time it takes for all correct nodes to deliver a broadcast message).
This allows us to make a simple but robust analysis that justifies the informal observations that have been made on these algorithms. This can be useful, for instance, to determine the appropriate algorithm for a given use case.
Also, this work opens research opportunities for protocols achieving yet unexplored latency/communication trade-offs, as well as the potential for interesting impossibility results.
Additionally, we are able to make a more robust argumentation on deciding whether an algorithm can be deemed as suited for the permissionless setting, \ie its performances should scale with the number of participants.

% recall previous known results
\subsection{Related Work}
\label{sec:relwork}

To the best of our knowledge, the closest work to our is from Fitzi\etal~\cite{parallelChains}. Like us, they define performance-related metrics (namely, delay and throughput) that they evaluate for several protocols. However the main focus of their work is a technique to improve the performance of some blockchain protocols. As such, their tools are mostly used to asses their proposal in regards to a few bitcoin-like protocols, and our analysis is much more complete. Vukolic~\cite{POWvsBFT} gave a good informal overview on how Proof-of-Work (PoW) and the classical BFT paradigm relates to performance/scalability trade-offs, but only made high level observations without quantification. Although Cachin \& Vukoli\'c~\cite{BCInTheWild} reviewed several algorithms, their work focus on resiliency and fault tolerance and not performance. Gervais \etal~\cite{SecPerfPoW} analyzed how tuning the performance-related parameters of PoW blockchains impacts their security. In contrast, we analyze performance as a function of security parameters and network model. Dib \etal~\cite{IARAPaper} reviewed and compared the performance of several algorithms, but their work is focus on the permissioned setting, and displays practical numbers for the end user rather than asymptotic complexity. BLOCKBENCH~\cite{blockbench} is a framework from Dinh \etal to evaluate the practical performance of blockchain \emph{implementations}, that moreover targets only the permissioned setting.

Otherwise, as it is standard practice, several works provide a complexity analysis of their algorithms. However they cannot always be compared since they may be calculated for different metrics, and under different models. In fact, a non negligible part of our work was to cast (and check) the proper values into our framework. Moreover, except for a few mentioned cases, the respective authors of the selected protocols made only a partial complexity analysis, \eg only for one of the metrics.

\subsection{Paper Organization}

This Paper is organized as follows: In Section \ref{sec:bg}, we make a short description of the concepts relevant to our analysis. In Section \ref{sec:methodo}, we describe the adopted methodology in terms of parameters and metrics used to make the evaluation. In Section \ref{sec:results}, we detail the network behavior of each protocol and give the steps leading to the results. In Section \ref{sec:discussion},  we give a short analysis of the results and motivate our arguments. Finally, we conclude this paper in Section \ref{sec:conclu}.

\section{Background}\label{sec:bg}

\subsection{Notations}
We denote by $negl(\cdot)$ a negligible function, \ie  such that for every polynomial $p$, $\exists \kappa_0, \forall \kappa > \kappa_0, negl(\kappa) \leq \frac{1}{p(\kappa)}$.\emph{With overwhelming probability} is abbreviated \wop.$|\cdot|$ is the size operator, \ie the number of elements for a set, or the length for a tuple. We use standard definition of $\Omega(\cdot), \Theta(\cdot), O(\cdot)$ notations for complexity, \ie $O(f(n))$ is of order at most $f(n)$, $\Omega(f(n))$ is of order at least $f(n)$, $\Theta(f(n))$ is of order exactly $f(n)$. We will further clarify their usage in our context in Subsection \ref{sec:params}. We use the terms protocol and distributed algorithm interchangeably.

\subsection{Definitions}

A Total Order Broadcast protocol, also called Atomic Broadcast, is a distributed algorithm that implements two primitives, $\texttt{TO-broadcast}(m)$ and $\texttt{TO-deliver}(m)$ with $m$ some message. We say that a process $p$ TO-broadcast $m$ when it executes $\texttt{TO-broadcast}(m)$, and  that $p$ TO-delivers $m$ when it executes $\texttt{TO-deliver}(m)$. To be a Total Order Broadcast, the protocol has to satisfy the following properties~\cite{TOSurvey}:
\begin{itemize}
\item \emph{Validity}: If a correct process TO-broadcast a message $m$, then some correct process eventually TO-delivers $m$.
\item \emph{Agreement}: If a correct process TO-delivers a message $m$, then all correct processes eventually TO-delivers $m$.
\item \emph{Integrity}: If $p$ is a correct process, then for every TO-delivery of $m$ by $p$, $m$ has been TO-broadcast.
\item \emph{Total order}: If two correct processes TO-deliver two messages $m_1$ and $m_2$, then both processes TO-delivers the messages in the same order.
\end{itemize}

Such protocol can be used to implement State Machine Replication, although the latter generally refer to a context were external clients sends commands to a network of replicas, which are consistent states machines. However the differences are mostly related to application semantics, and thus both concepts have been used interchangeably.

The studied protocols does not all solve exactly the same problem. In particular, they do not make the same assumptions about the underlying network model, and do not achieve the same properties, although they are similar. To make a consistent analysis, as mentioned earlier, we consider them all as Total Order Broadcast protocols. We consider they implement the primitives $\texttt{To-Broadcast}(m)$ and notify messages delivery by calling $\texttt{TO-deliver}(m)$. We will further specify how these primitives are mapped to each protocol in order to solve total order Broadcast, and will make note on the model variations adopted by each protocol.
A notable difference specific to Bitcoin and PoW-based algorithms is the weakening of the safety properties, which are only guaranteed \wop, whereas more traditional algorithms may not even be probabilistic.

\section{Methodology}\label{sec:methodo}

This section first describe the common model adopted, through parameters that capture the relevant assumptions made by all protocols. Then we explain how we related them to the security parameter to make meaningful statements, and we finally define our metrics.

\begin{figure}
\label{fig:params}
\begin{mdframed}
  \caption{Summary of parameters}
  \begin{itemize}
  \item[ ] $n$ number of participants
  \item[ ] $\Delta$ bound on network delays
  \item[ ] $\alpha$ byzantine power
  \item[ ] $\kappa$ security parameter
  \end{itemize}
\end{mdframed}
\end{figure}

\subsection{Network Model}

\subsubsection{Common basis}

All protocols analyzed are based on the Interactive Turing machines~\cite{UCFramework} model. Further assumptions on the network are appropriately stated through additional parameters that are given to each protocol at its instantiation. Other unrelated assumptions are not made explicit through this analysis. Each algorithm may require additional independent parameters that modify its behaviour.

More precisely, the protocols assume a set $\Pi$ with $|\Pi| = n$ of processes (also called nodes) with asynchronous clocks that exchange messages through asynchronous reliable point-to-point channels. Asynchronous means that there is no bound on the number of time steps taken for a message to be sent and received. Reliable means that no messages are dropped or modified. The communication graph is bidirectional and complete \ie any node can send a message to any node.

\subsubsection{Additional Assumptions} 

Next, we define a few more parameters that convey the additional assumptions made on the network model by the different algorithms. However the reader should keep in mind that this does not fully capture all subtleties of the models considered by each protocol.

\paragraph{Synchrony} We denote by $\Delta$ the maximum number of time steps taken by a message to reach its destination. Strongly synchronous protocols require $\Delta$ in their parameters, or a bound thereof. Partially synchronous protocols, depending on how they are formulated, either: require $\Delta$ in their parameters, but will simply not terminate if the given value is lower that $\Delta$, or not require it at all. 
For asynchronous protocols, $\Delta$ is a unit for asynchronous rounds defined as follow: Messages are assigned with a round number $r$ such that all $r-1$ messages must be delivered before sending any $r+1$ message.

In any case, evaluating the performance is only relevant when the protocols terminate, hence we will ignore messages schedules that prevent termination.

It should be noted that the nodes bandwidth is only indirectly taken into account: the bounds on network delay are supposed to hold such that all messages can be fully transmitted. 

%Additionally, we will repeatedly use the standard notion of \emph{rounds}. In the synchronous case, all message sent at the beginning of a round are received at the end. In the asynchronous case,

\paragraph{Byzantine faults~\cite{BFTGenerals}} Furthermore, all protocols tolerate some level of byzantine failure, that is, nodes that behave arbitrarily. We say that a node is correct or honest if it follows the protocol. Otherwise, we interchangeably use the terms malicious, faulty or Byzantine.
To do so, all algorithms need some sense of a sybil resistant resource \ie, a quantity that the adversary cannot artificially increase. This is translated by the assumption that no more than a fraction $\alpha$ of the resource is owned by malicious nodes.
Therefore, we define a generic parameter $\alpha$, which denotes the fractional byzantine "power" owned by the adversary. More precisely, depending on each assumption:
\begin{itemize}
\item If there is at most $t$ byzantine nodes in $\Pi$, then $\alpha = \frac{t}{n}$
\item If PoW is assumed, then $\alpha$ is the fraction of the total hashrate that byzantine node can produce.
\item If nodes have stake, then $\alpha$ is the the fraction of the total stake that is owned by byzantine nodes.
\end{itemize}
The reader should keep in mind that although the adversary power is modeled only through a unique parameter, the practical means and difficulty to acquire a given $\alpha$ is vastly different depending on the case considered above, \eg it means something completely different to control $t$ specific computers rather than being able to produce $x$ hashrate. Moreover, further assumptions can be made to make the adversary more or less powerful, especially regarding its capacity to dynamically corrupt nodes. 
%In particular, the set of byzantine nodes may be assumed to stay constant or not through protocol execution, refereed to as a static or adaptive adversary, respectively. Also considered is the weakly adaptive case, which is adaptive but subject to some delay for node corruption. 

\subsection{Security \& protocol parameters}
\label{sec:params}

Moreover, all protocols require a parameter to tune the desired security level. This parameter will be treated separately in this paper, and hence will be specifically named $\kappa$. Then, \emph{with overwhelming probability} means with probability $1-negl(\kappa)$.

Then, note that the precise requirement for each protocol is of the form: "for all admissible model parameters $n, \Delta, \alpha$, all stated properties holds with probability $1-negl(\kappa)$". 

%In practice, the probability of failure will also depend on other complex variables that will satisfy this requirement themselves.

However, this statement only deals with probability of failure when $\kappa$ is varying and the model parameters are fixed. As a result, all $O(\cdot)$ notations used to prove only refer to $\kappa$ and treat the model parameters as constants. 

This is problematic for our purposes because we also want to express the performance variation when the model parameters are varying (and in particular, $n$). Hence, we cannot make the same usage of $O(\cdot)$ notations for our metrics\footnote{This would result mostly in $O(\kappa^c)$ values, for some $c\geq 0$}.

One solution could be forbid ourselves $O(\cdot)$ notations where the constants also depends on some model parameter.
However, it is common to have the performance metrics depend on some complex variable which is shown to be $O(\kappa)$, but its precise dependencies to the protocol parameters are difficult to track, or even unknown.

As a result, we choose a middle ground where $O(\cdot)$ notations may still depend on the model parameters, but we will note in subscript on \emph{which} of them it does depend. For instance, $\Omega_{\alpha,\Delta}(\kappa)$ is a variable of order at most $\kappa$, and that does not depend on $n$.

It should be noticed that this still hides bounded expressions of the model parameters, for instance, we have that $\alpha = O_\alpha(1) =  O(1)$ because $\alpha$ is bounded by a constant independent of $\alpha$, namely, $1$. In practice we will avoid such simplifications and keep the subscript.
%we have that neither $\Delta$ nor $\alpha^{-1}$ are $O(1)$.

%The result is expression mostly of the form $f(p_1,p_2)\Omega_{p_3}(g(\kappa))$

Interestingly, each protocol may not require to be instantiated with the same parameters for all nodes. This is the case for instance for $\kappa$ in Bitcoin, where each node decide how many confirmations are deemed secure. On the contrary, in Algorand $\kappa$ has to be uniquely chosen before the protocol starts.

\paragraph{Additional parameter}
Beyond the previous model parameters, each protocol can be instantiated with additional protocol parameters. Since there will be at most one such parameter relevant for our analysis for each protocol, we we will name it $p$ in the following. This parameter is may be used to make efficiency and/or security trade-offs, or its optimal value may be unknown. Their precise meaning will be detailed in their corresponding section.

\subsection{Metrics}

We analyze the standard metrics used to evaluate distributed algorithms efficiency: latency and communication complexity. These are even more relevant in our context because they are directly experienced by the end user.

\paragraph{Latency} is the number of time steps between a message TO-broadcast and its TO-delivery by the last honest node. Note that, assuming synchrony, one round is $\Delta$ time steps. In the context of cryptocurrencies, it is directly the time the end user has to wait before his transaction is effective.

\paragraph{Communication complexity} is the amortized number of bits received by honest players for a message to be TO-delivered by all honest players. It should be noticed that communication complexity is usually given in terms of overhead, but we take here the total value, to give a sense of the network capacity required to run the protocol.  
This metric will be given as a function of the total message size $b$. 
%Simultaneously taking into account the message size and making an amortized cost may seem contradictory at first. In our case, $b$ is needed to assess the communication costs independently of messages contents, ignoring effects due to \eg batching, transactions queues, \etc. The amortized cost however is simply taking this size-dependent cost averaged over repeated executions, in order to reflect more accurately the cost of continuous operation.
As a side effect, the use of $b$ allows informally to distinguish a "constant overhead" cost, \ie that occurs independently of the data to broadcast, from an "efficiency" cost which reflects the overhead that scales with the amount of data transmitted.
We also remark that, depending on the algorithm, the bound on network delays $\Delta$ may also cover messages of size $b$. Thus, in practice, increasing $b$ could in turn increase $\Delta$.

We will use the term confirmation time as a synonym to latency, and bit complexity as a synonym to communication complexity.
% and messages are either constant size or dominated by the entry size (which is also deemed constant, in this context).

%For both metrics, we give a bound on their actual value \ie a worst case complexity. In that sense, we only give the \emph{guarantees} provided.

\section{Analysis}\label{sec:results}

%Before stating our results, we point out that the value given for the message complexity is actually the \emph{per node} value, \ie, it has been divided by $n$. This is relevant, because it capture the intuition that broadcast serves as the basic communication primitive in consensus algorithms, \eg as it is the case with gossip protocols. Moreover, since it may be reasonably to assume that the total network bandwidth scales with the number of nodes (otherwise communication would effectively be hindered), measuring network usage on a per node basis is more useful to evaluate the protocol behavior.

%running time does not depend on lambda => transaction finality

We now present our complexity results. This section is divided in three subsections, corresponding to different kinds of algorithms and \textit{in fine} analysis methodology. The first subsection treats blockchain style algorithms through a common framework from Garay \etal ~\cite{GKL17}. The second subsection evaluates protocols that follows the existing line of research for Byzantine Fault Tolerant (BFT) algorithms. The last subsection gathers protocol that work differently from the others and required specific treatment.

\subsection{Using the Bitcoin backbone protocol}

For Bitcoin, Snow White, and Ouroboros we will rely on the backbone protocol formalism from Garay \etal ~\cite{GKL17}. These protocols follows the classical blockchain approach, where leaders periodically broadcast new blocks, resulting in a tree structure where a common chain can (hopefully) be extracted. In Bitcoin, protocol participants are miners.

Their goal is to satisfy three properties: $k_p$-Common Prefix, ($k_g$,$\tau$)-Chain Growth and ($k_q$,$\mu$)-Chain Quality, abbreviated CP, CG and CQ respectively. They can in turn be used to prove to prove $k$-Persistence and $u$-Liveness. As a side note, in the first version of the backbone protocol~\cite{GKL15}, a weaker version of CP was considered which did not allow for a black box reduction. Pass \& Shi~\cite{BackboneImprov} further introduced an additional property, future self-consistency, solving this issue. We refer here to the strong version of CP, that includes future self-consistency, as in the cited version of the backbone protocol.
In this framework, $k$-Persistence and $u$-Liveness are properties of a Robust Transaction Ledger, which implies State Machine Replication~\cite{hybrid}. The $k$ parameter is the number of confirmations that a block must get to be committed, and $u$ is the number of round that a transaction takes before being committed.

Informally, CP states that pruning $k_p$ block from a chain results in a prefix that is shared by all nodes. CG states that every $k_g$ consequent rounds each node adds at least $\tau$ blocks, \ie $\tau$ is a round/block conversion parameter. CQ states that in every chain segment of length $k_q$, there is at least a proportion $\mu$ of honest blocks.
The proof for persistence is straightforward from CP. However the proof for liveness yields a value from the confirmation time as a function of the properties parameters, namely, $u=max(k_g, \frac{k_q+k_p}{\tau})$. 

We sketch the general idea of this proof here: 
After $u \geq k_g$ rounds, there are at least $\tau u$ new blocks in the honest players chain (CG). Therefore, if $\tau u > k_q + k_p$, after $u$ rounds, the honest players' chain grew by at least $k_q + k_p$ blocks. In this new chain, the first segment of length $k_q$ is a common prefix to all honest players, because it is $k_p$ blocks deep (CP). Moreover, there is at least $k_q \mu \geq 1$ blocks in this segment that are from honest players, which contains the transaction. Thus it is reported as stable by all honest players. Therefore, if $u\geq k_g$ and $\tau u > k_q + k_p$, then the protocol satisfy $u$-Liveness.

With foresight, we have that $k_q + k_p$ will be a security parameter. Thus, we will additionally detail their value depending on the model parameters. 

For these protocols, the analysis regarding communication complexity is simpler: Each block of size $b$ is broadcast once, and subsequent required confirmations does not change the amortized cost since each of them will (eventually) commit $b$ additional bits. The minimal communication cost is $b n $.
Furthermore, this cost is only increased by orphaned blocks. However, the properties of a robust ledger alone does not allow to bound the number of orphaned blocks. We observe that the three Bitcoin-Backbone based protocols satisfy an additional property: In $k_b$ rounds, there are at most $k_b \tau_b$ blocks produced.
Along with CG with $\tau > 0$, this bounds the ratio of orphaned blocks by $\Theta_\alpha(1)$.

%%%%%%%%%%%%%%%%%%%%%%%%%%%%%%%%%%%%%%%%%%%%%%%%%%%%%%%%%%%%%%%%%%%
\subsubsection{Bitcoin}~\cite{BTC}

For a baseline comparison, we start the analysis of Bitcoin. We build our analysis on the work from Garay \etal ~\cite{GKL17}, where they showed that Bitcoin satisfy the three properties Common Prefix, Chain Quality and Chain Growth.

In this work the authors articulate their proofs on the assumption of a \emph{typical execution}, that roughly states that parties produce blocks at a rate close enough to their expected value (\ie, hashpower). Then they show that any execution of $k$ rounds is typical with probability $1-e^{-\Theta_\epsilon(k)}$, where $\epsilon$ is a variable that quantify how close to its expected value is the block production. In turn, the proofs will rely on $\epsilon$ being "not too large", which is implicitly assumed in the "honest majority assumption". This assumption gives a bound on $\epsilon$ that depends on $\alpha$ and $\Delta/p$, where $p$ is the expected block creation time. Hence, an execution longer than $\Omega_{\alpha, \Delta/p}(\kappa)$ rounds is typical \wop.

CP, CQ and CG are proven assuming a typical execution, with $k_p = k_q \geq 2 k f$ and $\tau = (1-\epsilon)f$ for some variable $f$, hence we have $u$-Liveness with $u \geq \frac{4 k}{1-\epsilon} = \Theta_{\alpha, \Delta/p}(\kappa)$.

%The formulation of the the following protocols is a bit different. They satisfy the three properties unconditionally, \ie for all parameters $k_p, k_g, k_q \in \mathbb{N}$. However, the fact that the probability of failure is negligible is embed in the property itself. As a result, this formulation explicitly require that each $k_p$, $k_q$, $k_g$ is $\Omega(\kappa)$. Therefore, our methodology for them is to 

%%%%%%%%%%%%%%%%%%%%%%%%%%%%%%%%%%%%%%%%%%%%%%%%%%%%%%%%%%%%%%%%%%%
\subsubsection{Ouroboros}~\cite{Ouroboros}

Ouroboros is a strongly synchronous Proof-of-Stake (PoS) protocol. It is currently implemented for cryptocurrency Cardano (\url{https://www.cardano.org}).

Ouroboros does not assume PoW, and consequently has to generate its own randomness for the leader election. This is done inductively by leveraging previous broadcast to implement a common coin tossing. This part is somewhat complex but fortunately our analysis is blind to its details. Moreover, it gives the useful property of having one unique designated leader per round.

As a result, there is at most one block per round, and CP, CG and CQ can be proven using solely properties of the distribution of honest and malicious leader in the block tree.
This is done through the analysis of characteristics strings, which are an encoding of the schedule of honest and malicious leaders. Then they show that a string of length $k$ is forkable with probability $negl(k)$. This is, of course, for an appropriate notion of fork that is latter used to prove CP CQ and CG. In particular, since the distribution of characteristics strings only depends on $\alpha$, we have that $k = \Omega_\alpha(\kappa)$ 
Moreover, $\tau = 1-\alpha$, because there is at least one block for each round with an honest leader.

As with Bitcoin, strong synchrony is assumed, therefore the latency is $\Delta\Omega_\alpha(\kappa)$

%%%%%%%%%%%%%%%%%%%%%%%%%%%%%%%%%%%%%%%%%%%%%%%%%%%%%%%%%%%%%%%%%%%
\subsubsection{Snow White}~\cite{SnowWhite}

Like Ouroboros, Snow White is a strongly synchronous protocol that does not assume PoW. Its particularity however is its tolerance of honest nodes that do not show up.

Snow White is able to cope with honest nodes being asleep, as long as there are more awake honest nodes than malicious ones. In contrast, asleep honest nodes in Bitcoin are counted as malicious, at least until they are implicitly dropped during difficulty adjustment. In that sense Snow White tolerates stronger adversaries, because for $\alpha$ malicious nodes, it requires only $\alpha$ honest ones and $1-2\alpha$ asleep ones instead of just $1-\alpha$ honest nodes.

The notations used are slightly different, because all time-related values are directly expressed in terms of number of time steps instead of number of rounds. As a result, the Chain Growth $\tau$ parameter directly translate blocks to time steps, and we can simply rely on $k_g+k_p$ being $O(\kappa)$.

Snow White works somewhat similarly to Ouroboros, in the sense that it inductively generates its own public randomness to elect leaders. However, each node has a fixed probability $p$ to be elected at each time step, instead of having unique leaders at each round. The authors then prove Chain Growth with $\tau^{-1} = O_{\alpha, n p}(1) + \Delta O(1)$. Roughly speaking, the first term is the expected time it takes for an online node to find a block, and the second one is the time for the block to be sent. Thus we have a total latency of $\Delta \Omega(\kappa) + \Omega_{\alpha,n p}(\kappa)$
% $\Omega(\Delta\lambda + \lambda_{\alpha,n p})$.

\subsection{Traditional BFT style algorithms}
The following algorithms are more inspired from the existing literature on BFT consensus algorithms, and share some common concepts that we outline here for convenience. First, they all follow the framework of a per-block consensus instance, which is final and essentially independent from the others. This instance is implemented as a procedure that each node starts with its own input, and then the output is the next block to be added.

These protocols often make use of simpler common primitives. One of them is Reliable BroadCast (abbreviated RBC), that let a designated sender to broadcast a value such that honest nodes either agree on the value sent, or that the broadcast was unsuccessful. Formally, a Reliable Broadcast protocol is exactly TO-Broadcast, without the Total Order property~\cite{BFTGenerals}.

Another primitive used is the Binary Agreement (abbreviated BA) primitive. In a BA, each processor propose its own input value $v_i \in \{0,1\}$, and deliver another value, such that:
\begin{itemize}
\item \emph{Agreement}: If a correct processes delivers a value $v$, then all correct processes delivers $v$.
\item \emph{Validity}: If a correct process delivers $v$, then $v$ was proposed by a correct process.
\item \emph{Termination}: All correct processes eventually decides.
\end{itemize}

%In general, the analysis for these protocols is simpler, and the authors already include

%With this framework, it is always possible to 

%%%%%%%%%%%%%%%%%%%%%%%%%%%%%%%%%%%%%%%%%%%%%%%%%%%%%%%%%%%%%%%%%%%
\subsubsection{Tendermint}~\cite{tendermintthesis}

Tendermint is a partially synchronous TO Broadcast algorithm, targeting high performance in a small-scale permissioned setting.

Its normal case operation is reminiscent of PBFT~\cite{PBFT}, except that leaders are changed at each command (\ie block). For each block, there is a leader that will execute RBC implemented through two all-to-all voting rounds. Of course, to optimize bandwidth, only the hash of the block is included in the voting messages. This require the $b$ bytes of the batch to be sent to all $n$ nodes, and then the Reliable Broadcast cost of $n^2$ bytes, taking three of communication steps. That is, in a fault-free execution, $(bn+n^2)\Theta(1)$ bit complexity and $\Delta \Theta(1)$ latency. 

Note that Tendermint, as PBFT, is optimized for the normal case operation.
However, since the leader is arbitrarily designated, it may be malicious. In this case, the properties of the Reliable broadcast still ensure safety, but termination can be prevented. This is why, after a timeout, it will be the turn of the next designated leader to proceed for a new broadcast. The choice for the new leader being in a fixed round-robin order, all $\alpha n = \Theta_\alpha(n)$ malicious nodes may be leaders first and delay the block commit by the same factor. Therefore, the worst case communication cost is $(bn^2+n^3)\Theta_\alpha(1)$ bits and $\Delta n \Theta_\alpha(1)$ latency.

%%%%%%%%%%%%%%%%%%%%%%%%%%%%%%%%%%%%%%%%%%%%%%%%%%%%%%%%%%%%%%%%%%%
\subsubsection{HoneyBadgerBFT}~\cite{HBBFT}
HoneyBadgerBFT (abbreviated HBBFT) is an asynchronous probabilistic algorithm that solves TO-broadcast.

It is built on the Asynchronous Common Subset (ACS) primitive. Informally, ACS is used to agreed on a subset of the union of the nodes' inputs. It is defined as follow~\cite{HBBFT}:
\begin{itemize}
\item \emph{Validity}: If a correct node deliver a set $v$, then $|v| \geq (1-\alpha)n$ and $v$ contains the inputs of at least $(1-2\alpha)n$ correct nodes.
\item \emph{Agreement}: If a correct node deliver $v$, then every node deliver $v$.
\item \emph{Termination}: All correct nodes eventually delivers.
\end{itemize}

To prevent the adversary from selectively censoring transactions included in the output set, HBBFT uses a threshold encryption scheme before broadcasting transactions.

The ACS is a reduction to Reliable Broadcast and binary Byzantine Agreement from Ben-Or \etal ~\cite{ACSReduc}.
Roughly, each node runs in parallel an instance of RBC, to send their value. Then, for each RBC, there is a BA instance to agree whether it did terminate. Then, the nodes outputs the union of RBC values for which its termination has been agreed. To ensure the Validity property, the protocol has to ensure that at least $(1-\alpha)n$ RBC instances terminated.

The RBC terminates in three rounds and has $|m| n O(1) + n^2 log(n)O(\kappa)$ bit complexity, with $|m|$ the message size. The BA is an asynchronous probabilistic protocol with $ n^2 O(\kappa)$ bit complexity. At each round it has $1/2$ probability to terminate. As stated above, HBBFT has to wait for $(1-\alpha)n = n \Theta_\alpha(1)$ parallel instances of them to terminate, the time for this to happen is $log(n) O_\alpha(1)$ rounds.

%The authors of HBBFT provide a detailed complexity analysis, however th
However one of HBBFT achievements is that at the end of this procedure, it commits data from all nodes input. That is, if all nodes have an input of size $b$, then the batch committed will be of size $n b \Theta(1)$ bits.
Hence, at each batch of size $b$ committed, there is $O(1) bn + n^3 log(n) O(\kappa)$ bits received by honest players. 

Note that the authors do provide an analysis, but they state their results in terms of \emph{overhead}, \ie the total cost divided by $b$. Furthermore, by specifying a batching policy $b = n^2 log(n) O(\kappa)$, they obtain the $n O(1)$ figure which is a constant per-node overhead. This emphasis on the low overhead is less visible in our results, although it is translated by the fact that HBBFT complexity on the $b$ factor is $b n$ instead of $b n^2$ for other BFT-style protocols, \ie a reduction factor of $n$ consistent with the authors analysis.

\subsection{Others Blockchain based algorithms}

The two following protocols adopts significantly different paradigms, such that they are treated separately.

%%%%%%%%%%%%%%%%%%%%%%%%%%%%%%%%%%%%%%%%%%%%%%%%%%%%%%%%%%%%%%%%%%%
\subsubsection{Spectre}~\cite{Spectre}

Spectre is a PoW-based partially synchronous distributed algorithm aimed specifically at a cryptocurrency application.

In Spectre, the PoW merely serves as a network-level synchronization primitive. As such, there is little restriction on the mining process, and the mining hardness $p$ is only a protocol parameter for tuning network behavior. More precisely, the only assumption regarding $p$ is that blocks are produced less quickly that the nodes can receive them, hence it is \emph{not} related to $\kappa$.

The Spectre protocol maintains a public ledger that grows over time, and a sub-protocol $RobustTxO$ outputs the linearized log with the safety and liveness properties.
Once again, participants to the protocol are miners. A message is a transaction, and we execute TO-Broadcast when including it in a block. A transaction is TO-Delivered when it is output by $RobustTxO$.
Then, under some conditions that we detail below, Spectre is a (variant of a) TO Broadcast, because~\cite{Spectre}: Proposition 5 (Safety) implies Agreement and validity, Proposition 4 (Consistency) implies Total Order, and Proposition 6 (Weak Liveness) implies Termination.

Because it is targeted for a cryptocurrency application, Spectre achieves a weaker version of TO-Broadcast in two regards. First, the total order is weakened into a partial order, meaning that the messages are only ordered if the application requires so. Secondly it has a weakened liveness property. Translated to the TO-Broadcast abstraction, it means that Termination may not be satisfied under some conditions on the nodes' inputs. This weakening is acceptable for Spectre because those conditions express nodes honesty according to the application context, \ie honest cryptocurrency users that never try to double-spend will enjoy termination.

Roughly speaking, transactions are extracted from the blocks graph, and it is possible for each of them to compute whether their probability to be undone is greater than a given $\epsilon$. Then, the $\Omega_{\alpha}(\kappa)$ confirmations are still required, but since block creation is more flexible, the $\delta$ is no longer a multiplicative factor but an additive overhead that ensure a consistent view across all honest nodes.

More precisely, Spectre implements a procedure $RiskAcceptTx$ that gives a bound on the probability that safety will not hold for this transaction. The authors then show than the bound returned is smaller than a given $\epsilon$ after $log(\frac{1}{\epsilon})O(1)$ honest blocks are created. However, since this procedure additionally require the node local values for $\alpha$ and $\Delta$,  we have that the number of honest blocks required is $O_{\alpha, \Delta}(\kappa)$. The time taken for this to happen is obtained by multiplying it by $\frac{1}{(1-\alpha)p} = (p \Theta_\alpha(1))^{-1}$. Then for all nodes to be aware of these blocks an additional $\Delta$ overhead is added, resulting in $\Delta O(1)+\frac{O_{\alpha, \Delta}(\kappa)}{p}$ latency.

Communication complexity on the other hand is simpler to analyze. In fact, the analysis is similar to the one for others blockchain style algorithms. Since Spectre selectively outputs transactions individually, the adversary only mean to increase the communication complexity is to use all its hashpower to send invalid transactions, effectively increasing the amortized cost by a factor $\frac{1}{1-\alpha} = \Theta_\alpha(1)$.

%%%%%%%%%%%%%%%%%%%%%%%%%%%%%%%%%%%%%%%%%%%%%%%%%%%%%%%%%%%%%%%%%%%
\subsubsection{Algorand} ~\cite{Algorand}, ~\cite{AlgorandBA}

Algorand is a strongly synchronous protocol with a per-block agreement approach.

Its byzantine agreement primitive vastly differs from traditional BFT algorithms. In fact, like Ouroboros and Snow White, Algorand rely on a public randomness computed in previous blocks. It is used to elect a committee (instead of a leader) \emph{at each round} that will have sufficiently many honest nodes, \wop. These committees run a byzantine agreement which does not require a private state from the nodes (except from their private key), since committees would not be able to pass it on to the next committee properly. This property is called \emph{player replaceability} by the authors. Except from this property, common techniques from BFT algorithms can be used to reach agreement in a constant number of rounds. In particular, since it has been ensured that committees have a constant fraction of honest nodes, standard quorum-based arguments are still valid.

More precisely, at each round, each user has a fixed probability $p$ to be part of the committee. To bound the number of malicious nodes in a committee, the authors leverage the fact that $\forall \beta < \alpha$, the probability of having at most $\beta k$ malicious nodes in a uniformly sampled committee of size $k$ is $1-negl(k)$. Thus, the committees expected size is $\Omega_{\alpha}(\kappa)$.

As a result, each round that would be equivalent to a $n^2$ all-to-all communication in a traditional BFT algorithm is now a "committee-to-all" $n \Omega_{\alpha}(\kappa)$ communication.

To optimize bandwidth, Algorand still does some sort of leader election. More precisely, the expected committee size for the block proposers is the smallest such that there is at least one proposer \wop, that is $\Theta(\kappa)$.

As a result, Algorand's communication complexity is $b n\Theta(\kappa) + n \Omega_{\alpha}(\kappa)$ and its latency $\Delta O(1)$.

\section{Discussion}\label{sec:discussion}

\begin{table}[tbp]
\centering
\caption{Performance summary}
\label{tab:perfs}
\begin{tabular}{@{}lll@{}}
\toprule
Algorithm                          & Latency                                    & Communication complexity           \\ 
\midrule
Nakamoto~\cite{GKL17}              & $\Delta\Theta_{\alpha, \Delta/p}(\kappa)$  & $b n \Theta_\alpha(1)$                              \\
Ouroboros~\cite{Ouroboros}         & $\Delta\Omega_\alpha(\kappa)$              & $b n \Theta_\alpha(1)$                              \\
SnowWhite~\cite{SnowWhite}         & $\Delta \Omega(\kappa) + \Omega_{\alpha,n p}(\kappa)$     & $b n \Theta_\alpha(1)$                              \\
Spectre~\cite{Spectre}             & $\Delta O(1)+\frac{O_{\alpha, \Delta}(\kappa)}{p}$ & $b n \Theta_\alpha(1)$                              \\
Algorand~\cite{Algorand}           & $\Delta O(1)$                                   & $b n\Theta(\kappa) + n \Omega_{\alpha}(\kappa)$ \\
Tendermint~\cite{tendermintthesis} & $\Delta n \Theta_\alpha(1)$                & $(bn^2+n^3)\Theta_\alpha(1)$                      \\
HoneyBadgerBFT~\cite{HBBFT}        & $\Delta log(n) O_\alpha(1)$                & $bn O(1) + n^3 log(n) O(\kappa)$             \\
%RedBelly          & $\delta$                                   & $b n^2+n^3$                        \\ 
\bottomrule
\end{tabular}

\end{table}

Table \ref{tab:perfs} summarizes the different results for the studied protocols. These clearly outline the different styles of algorithms that required different methodologies in Section \ref{sec:results}. Pure Blockchain style algorithms are the most efficient regarding communication complexity. Their cost is even optimal in a sense, because $b n$ bits are required to make the messages known, and some (worst case) overhead due to byzantine nodes seems unavoidable. On the other hand, we can see that their latency suffers from their security parameter, which itself may be an unbounded function of the network parameters. 

We see that Algorand stands out in this table, because it is the only one whose latency does not depend on $\kappa$, but their bit complexity does. This is easily understandable in the fact that blocks in Algorand are final and that no fork can occur. 
Intuitively, we can see Algorand as doing all the random picking of participants in parallel, at the same time, instead of sequentially.

%We see that for Snow White, the overhead induced by the tolerance of sleepy participants is the dependence on $\Delta$ for their security parameter in the latency. 
%It can be noticed that Ouroboros showed this to be unnecessary in a setting with online honest users. 
%This can be explained by the necessity to run the leader election concurrently, similarly to PoW-based elections (although they do not assume PoW), which makes network synchronization intervene in the number of blocks required for confirmation.

\paragraph{Blockchain communication complexity}
It may be surprising that blockchain protocols have such low communication cost compared to traditional $O(n^3)$ algorithms. This can be explained by the fact that blockchain style algorithms share a common component: public, unbiased randomness. Note that, for instance, the hash of a block is indeed a public random, but not unbiasable, because the adversary is able to contribute to the input of the hash, thus allowing her to make the random value suits her needs.
This component is needed to leverage an important mechanism of blockchain protocols: Within a random sample of the protocol participants, each node has a probability $\alpha < 1$ of being malicious. Therefore, by repeatedly taking nodes at random, the total number of malicious nodes can be bounded \wop. This way, a finite number of unreliable broadcasts can be sufficient to reach agreement with high probability, hence the probabilistic safety.

\paragraph{Impact on the meaning of permissioned/permissionless}
Permissioned and permissionless are terms that are well understood informally, but their precise meaning can be problematic. Indeed, even if "being able to join the system at any moment" is somewhat clear, deciding whether a given algorithm is permissionless is arguably arbitrary. This is due to a possible argument that a simple transformation could relax any permissioned algorithm into a permissionless one. Indeed, in a permissioned setting, nodes are determined and can be given arbitrary weights (\ie having nodes count as multiple ones), thus translating the fault tolerance to a fraction of the total weight. Then, defining the weights as the nodes stake recorded in the blockchain itself, we should obtain a legitimate PoS protocol. This transformation has already been used as an argument in practice ~\cite{hashgraph}. Moreover, the inverse transformation (from permissionless to permissioned) being also trivial, it would seem that the distinction permissioned/permissionless would be more specific to the usage of the algorithm rather than the algorithm itself. However, this seems to be false in practice, as algorithms are \textit{de facto} targeted towards a specific setting.

Our work allow for an answer to this issue. In our sense, we say that an algorithm is suitable for the permissionless setting if it is scalable and stay efficient even with a very high number of participants. In our opinion, it should have a bit complexity linear in $n$, and a latency independent from $n$.
We notice that Pass \& Shi ~\cite{thunderella} proposed a definition from permissionless environments, in which they formally define a model were executions are permissionless/permissioned. In that sense, our statement is not a \emph{definition} of a permissionless environment in itself, but a criteria for categorizing algorithms as suited to the permissionless setting or not.

\section{Conclusion}\label{sec:conclu}
In this work, a wide range of popular blockchain protocols was analyzed. Our metrics and methodology has been carefully tuned to accurately describe the behavior of each protocol.

We showed that, when considered as variants of Total Order Broadcast protocols, permissionless blockchain protocols had a very efficient network usage, but were suffering latency-wise from their security parameter.
%We also notice than Algorand is the only protocol that explored different trade-offs compatible with a permissionless usage.
This work also can be used as a ground methodology to assess new protocols performances. 
Finally we discussed some impacts of these results on the classification of permissioned and permissionless algorithms.

%%%%%%%%%%%%%%%%%%%%%%%%%%%%%%%%%%%%%%%%%%%%%%%%%%%%%%%%%%%%%%%%%%%%%%%%%%%%%%%%

\section*{ACKNOWLEDGMENTS}
We are thankful to Petr Kuznetsov, Guillaume Hebert, Thomas Domingos, and Vincent Coquet for their help in fruitful discussions. This work was carried as part of the Blockchain Advanced Research \& Technologies (BART) Initiative. We are also thankful to the Technical University of Munich (TUM). This research work has been carried out under the leadership of the Institute for Technological Research SystemX, and therefore granted with public funds within the scope of the French Program \textit{Investissements d'Avenir}.

%%%%%%%%%%%%%%%%%%%%%%%%%%%%%%%%%%%%%%%%%%%%%%%%%%%%%%%%%%%%%%%%%%%%%%%%%%%%%%%%

\bibliographystyle{splncs04}
\bibliography{biblio}

\end{document}